\documentclass[aps,prb,floatfix,epsfig,twocolumn,showpacs,preprintnumbers]{revtex4-1}

\usepackage{amsfonts}
\usepackage{amssymb}
\usepackage{amsmath}
\usepackage{graphicx}
\usepackage{color}
\usepackage{hhline}
\usepackage{hyperref}
\hypersetup{colorlinks}
\hypersetup{
    linkcolor=blue,          
    citecolor=magenta,
    filecolor=magenta,      
    urlcolor=blue}

\begin{document}

\title{Electronic structure and thermoelectric properties of CoAsSb with post-DFT approaches}
\author{Andrey L. Kutepov,$^{1}$\footnote{e-mail: akutepov@bnl.gov} Anthony Ruth$^{2}$}

\affiliation{$^{1}$ Brookhaven National Laboratory, Upton, NY 11973-5000, USA}

\affiliation{$^{2}$ Department of Physics, University of Notre Dame, South Bend, IN 
60439}

\begin{abstract}
We study the electronic structure and thermoelectric properties of recently synthesized CoAsSb. The calculated bandgap becomes more accurate for increasingly-complex electronic structure methods: generalized gradient approximation, hybrid functionals, self-consistent linearized quasiparticle GW method (LQSGW), and LQSGW with simplified vertex corrections. By equating the bandgaps of each method from a rigid shift of the bands, we evaluate the contributions made to thermoelectric properties beyond the bandgap. In doing so, we evaluate the efficacy of a common-practice: a rigid shift applied to less-costly electronic structure methods in order to gain some accuracy of the more-costly methods. We find that while the shift made the Seebeck coefficients much closer to one another than from the original electronic structures, there remain differences in the Goldsmid-Sharp (thermoelectric) bandgap between the methods and from the intended electronic bandgap. Additionally, some lasting differences in temperature dependence remain between the methods. 
\end{abstract}

\maketitle

\section*{Introduction}
\label{intro}
Thermoelectrics, solid state heat enginges which generate a voltage from an applied temperature difference or vice versa, offer enticing advantages for a wide range of technological applications.\cite{twaha_zhu_yan_li_2016} However, sub-optimal performance characteristics of active materials have limited their use in power generation. The most important characteristic is the Seebeck coefficient (S) which is the ratio between the voltage and temperature gradients $\nabla V=S\nabla T$. When used in a circuit, current flows through the thermoelectric material, and the lower the electrical resistance the greater the power which can be drawn. This is reflected by the material's power factor, $S^2\sigma$, where $\sigma$ is the electrical conductivity. Finally, the power conversion efficiency can be found by dividing the electrical power by the heat flow, and this is typically expressed as a dimensionless figure of merit: $zT=\frac{S^2\sigma}{\kappa}$ where $\kappa$ is the thermal conductivity. $\kappa$ is composed of electronic thermal conductivity $\kappa_e$ and lattice thermal conductivity $\kappa_L$, $\kappa=\kappa_e+\kappa_L$. Finally, the power factor and figure of merit change drastically with temperature and a wide temperature range is used during power conversion, from ambient temperature to high temperature. Compared to other methods of generating electricity, thermoelectrics typically have low thermal conversion efficiency (low zT) but high power density (high $S^2\sigma$). Despite the drawbacks, thermoelectric power generation is still used in applications where the reliability of solid state devices is crucial such as electricity generation in remote locations.\cite{fihelly_baxter_1970} To date, the most widespread use of thermoelectrics is cooling where sub-optimal zT values do not as heavily affect device efficiency, power densities can be quite high, a compact form factor aids insulation, and a narrower temperature range is used.\cite{zhao_tan_2014}

The list of known materials with high thermoelectric efficiency, such as Bi$_{2}$Te$_{3}$,\cite{bjap_5_386} PbTe,\cite{science_321_554} Mg$_{2}B^{IV}$ ($B^{IV}$=Si, Ge, Sn)\cite{scrmat_69_606} and others is constantly increasing. The discovery of new thermoelectric materials has historically been experimentally-driven. The small bandgap of many thermoelectric materials is of the same order as the thermal energy (kT) especially at high temperatures, and this in turn is responsible for the large swings in device performance with temperature. Small bandgap semiconductors typically have the highest figure of merit because the Seebeck coefficient tends to increase with the bandgap,\cite{jem_28_869} but both $\sigma$ and $\kappa_e$ decreases.  Small bandgap semiconductors are in the regime where $\kappa_e$ and $\kappa_L$ are comparable. Materials with large unit cells are also ideal because $\kappa_L$ is minimized. A material which is limited by lattice thermal conductivity can be nanostructured to increase phonon scattering and reduce $\kappa_L$. The lattice thermal conductivity can be difficult to evaluate from first principles, but can be measured accurately by experiment. Despite the fact that experimentally-measured performance characteristics are more reliable than predicted characteristics, fast measurement cycles may not realize a material's full potential because optimal performance typically requires careful tuning through doping. 

The demand for better thermoelectric materials and the stringent requirements on material properties has made the use of high-throughput calculations to screen potential materials quite appealing. A very good indication of the importance of the subject is a recent publication of high-throughput calculations\cite{jmcc_4_4414} of thermoelectric properties for more than 48000 inorganic compounds from the Material Project. From the theoretical point of view, the accurate prediction of thermoelectric properties still represents a formidable task.\cite{thermoelectric_review} Not only are many complicated physical processes involved (electronic correlations, phonons, electron-phonon interaction) but also thermoelectric materials have sufficiently large unit cells (often containing about 10 atoms or more), that the application of advanced electronic structure methods is too costly in terms of computer time/memory. As a result, the majority of theoretical works use density functional theory (DFT) to get insight on the basic electronic structure of thermoelectric materials. In this respect, a good example is the above mentioned high-throughput calculations performed in Ref. [\onlinecite{jmcc_4_4414}], where the combination of DFT (as implemented in the VASP code\cite{prb_54_11169}) and the semi-classical Boltzmann equation (as implemented in the code BoltzTrap\cite{cpc_175_67}) was applied to calculate transport properties from electronic structure. Namely, authors of the Ref. [\onlinecite{jmcc_4_4414}] have compared the calculated Seebeck coefficient (S) and the thermoelectric power factor $S^{2}\sigma$. As it follows from Ref. [\onlinecite{jmcc_4_4414}], the performance of the combination DFT-band-structure plus Boltzmann equation is qualitatively good for the Seebeck coefficient with the primary source of discrepancy being the DFT underestimation of the bandgap. DFT underestimation of the bandgap is the leading error in both S and $S^{2}\sigma$, and naturally, is the principal target for adjustments. The accuracy of the power factor is worse because evaluation of the conductivity involves an additional approximation, namely a constant and universal relaxation time $\tau$.  As it follows from the Ref. [\onlinecite{jmcc_4_4414}], the results, especially for the Seebeck coefficient, can be considerably improved if one uses the "scissor" operator, i.e. rigid shift of calculated unoccupied bands relative to the occupied bands making the bandgap equal to the experimental one. The scissor operator corrects the bandgap, but leaves the band dispersion and corresponding effective masses unchanged.  It is, thus, interesting to see how an application of advanced electronic structure methods (which can address the bandgap issue more directly than DFT, avoiding the "scissor" operator) works in terms of the evaluation of S. 

Application of many body methods to the materials of thermoelectric importance is still rather rare,\cite{prb_88_045206,prb_93_205442,prb_82_085104,prb_90_075105,prb_88_165135,prb_82_245203} and it is limited to the simplest approach, namely one-shot GW (G$_{0}$W$_{0}$). In this approach, Green's function G and screened interaction W are evaluated based on DFT one-electron spectra and then the correction to these spectra is evaluated based on the first order self energy $\Sigma=G_{0}W_{0}$. Often additional approximations are involved, such as the plasmon pole approximation, diagonal only form of the self energy, and perturbative (first order) solution of the Dyson's equation. As it was shown in Refs. [\onlinecite{prb_90_075105,prb_93_205442,prb_88_045206}], the diagonal approximation for the self energy, often associated with G$_{0}$W$_{0}$ studies, fails in the case of Bi$_{2}$Tl$_{3}$, suggesting that wave functions (which usually are considered at the DFT level in the above approximation) undergo essential changes and one has to consider off-diagonal elements of the self energy in order to obtain a meaningful result. In this respect one can speculate that this actually might be true for thermoelectric materials in general, because as a rule, they possess small (0.1-0.3eV) bandgaps and DFT usually is unable to reproduce this bandgap, resulting in a metallic spectrum. The transition from a metal (in DFT approximation) to a semiconductor (in GW based methods) may be accompanied with a reconstruction of wave functions, which should not be neglected. Developing this idea, it is natural to assume that the effects of self-consistency might be more essential in this class of materials as compared to the wide bandgap semiconductors (such as Si, GaAs, ...). Semiconductors like Si are semiconductors already at the DFT level. GW approaches corrects the bandgap without a noticeable change in the wave functions. In thermoelectric materials, on the other hand, strong restructuring in wave functions discovered in a one shot study, might continue during subsequent iterations. This revelation makes the application of self consistent GW-based methods appealing.

Additionally, the small bandgap of thermoelectric materials can cause breakdown of another widely used approximation. Namely, the temperature dependence of transport properties. Frequently, the temperature-dependence of transport quantities is considered as coming entirely from the change in carrier concentrations with temperature, however, the chemical potential can shift as well and the former approximation would implying charging of the system. For small bandgap systems this approximation breaks down.

 Temperature-dependent electrical conductivity  of a recently synthesized\cite{cm_30_4207} material CoAsSb matched an activated band insulator with bandgaps derived from activation energies. However there are transitions between two different bandgaps. From 325-475 K, the bandgap was 0.124 eV whereas from 575-975 K the bandgap was 0.256 eV. The peak Seebeck coefficient (-134 $\mu V/K$) occurs in the former range at 450 K. The Goldsmid-Sharp bandgap (GS gap) $E_{GS}=2*S_{max}*T_{max}$ is 0.12 eV. The actual composition which was prepared and explored experimentally was CoAsSb$_{0.883}$. 

Theoretical study of this material \cite{cm_30_4207} consisted in the evaluation of the electronic structure in the local density approximation (LDA) with subsequent application of BoltzTrap to calculate transport properties assuming fixed chemical potential. LDA resulted in a metallic band structure contradicting with experimental results.\cite{cm_30_4207} Consequently there was a mismatch of the calculated transport properties in accordance with general tendencies discovered in high-throughput calculations.\cite{jmcc_4_4414} To compare ways of resolving the bandgap issue, authors of the Ref. [\onlinecite{cm_30_4207}] applied the modified Becke-Johnson functional mBJ (of semiempirical nature) and obtained a small bandgap of 0.074eV for the LDA-optimized geometry. For the experimentally-measured geometry, even with the mBJ functional, the band structure was gapless. The authors speculated that the presence of a bandgap in experiments can be explained by the influence of the defects. 

Thus, a natural continuation of the study of thermoelectric properties of CoAsSb and of the metal to small bandgap semiconductor problem is to calculate the electronic structure of CoAsSb with the use of self-consistent ab-initio many body techniques. In this work, we apply self consistent linearized quasi-particle GW method LQSGW\cite{prb_85_155129,cpc_219_407} to study electronic structure and thermoelectric properties of CoAsSb. We compare the LQSGW calculations with semiempirical hybrid functional PBE0\cite{jcp_105_9982} with different admix of Hartree-Fock non-local exchange (0.1, 0.15, and standard 0.25). In the second part of our study, we apply semiclassical Boltzman transport theory\cite{Boltzman} to evaluate thermoelectric properties of CoAsSb including temperature-dependence of the chemical potential. In this study calculations were performed using both experimental and optimized geometries from Ref. [\onlinecite{cm_30_4207}]. We have also checked the optimized geometry by performing independent optimization using VASP code and concluded that the structure obtained is essentially identical to the one obtained with Wien2k.

The plan of the paper is as follows. Section \ref{meth} discusses the methods for the calculations. Section \ref{res} provides the results obtained and the discussion. The conclusions are given thereafter.

\section{Methods}
\label{meth}

Electronic structure of CoAsSb was evaluated using the FlapwMBPT software package\cite{flapwmbpt} for the experimentally-measured geometry. FlapwMBPT implements diagrammatic approaches based on Full potential Linearized Augmented Plane Wave supplemented with Local Orbitals (FLAPW+LO). For this study, four approximations were selected. (1) the Generalized Gradient Approximation (GGA) as parameterized in PBE form.\cite{prl_77_3865} (2) The self-consistent linearized quasi-particle GW method, which was first introduced in [\onlinecite{prb_85_155129}], and additional details about the approach were given in [\onlinecite{cpc_219_407}]. (3) The modified LQSGW approach wherein W is evaluated only once based on time dependent density functional theory TD-DFT.\cite{prl_52_997} (4) Hybrid functional PBE0,\cite{jcp_105_9982} as well as variations of it with different percentage of the exact exchange. 

To be specific, in the modified LQSGW approach (3) W is evaluated from the reducible polarizability $\chi$ ($W=V+V\chi V$, where $V$ is the bare Coulomb interaction). $\chi$ is evaluated as the exact (in DFT approximation) functional derivative of the electronic density $\rho$ with respect to an external electric field $\phi$: $\chi=\frac{\delta \rho}{\delta \phi}$. When $\rho$ is calculated in DFT approximation, the evaluation of the functional derivative corresponds to the solution of the following equation:

\begin{equation}\label{chi}
\chi=\chi_{0}+\chi_{0}[V+\frac{\delta V_{xc}}{\delta \rho}]\chi,
\end{equation}
where $\chi_{0}$ is the non-interacting polarizability and $V_{xc}$ is the exchange-correlation potential of DFT. The polarizability $\chi$ (and correspondingly W), evaluated this way, mimics the vertex corrections and usually provides a good approximation, especially for metallic systems. It was demonstrated, for example, in the calculations of dynamical response functions in Na and Al.\cite{prb_84_075109} Thus, we believe it should be more accurate for small bandgap semiconductors than the corresponding quantity obtained in LQSGW. We will abbreviate the third approach as LQSGW$^{PBE}$ reflecting the fact that W is evaluated following its definition with PBE parameterization of the GGA approximation. Direct application of vertex-corrected GW schemes\cite{prb_94_155101,prb_95_195120} would be too time consuming to apply for CoAsSb. However, the above slightly more advanced method LQSGW$^{PBE}$ (as compared to LQSGW) still mimics true vertex-corrected GW calculations and improves the bandgap.

CoAsSb has a large unit cell (12 atoms), and its symmetry group (P2$_{1}$/c, arsenopyrite-type structure \cite{cm_30_4207}) has 4 operations. The evaluation of the band energies on sufficiently fine \textbf{k}-mesh in the Brillouin zone is necessary followed by subsequent interpolation to even finer \textbf{k}-mesh used to study transport properties. In this work, all electronic structure calculations were performed using $8\times 8\times 8$ \textbf{k}-mesh. For a 12 atom unit cell with an $8\times 8\times 8$ \textbf{k}-mesh, computational requirements of self-consistent LQSGW calculations were quite formidable. This issue was resolved by i) using real-space plus Matsubara's time implementation \cite{prb_85_155129} of polarizability and self-energy evaluation, which allows considerable time savings as compared to the traditional reciprocal-space plus Matsubara's frequency formulation and ii) extensively usage of Message Passing Interface (MPI) to distribute the computational workload. In this respect, we refer interested reader to our earlier publication\cite{cpc_219_407} where the details of our parallelization strategy for the evaluation of all principal ingredients of GW algorithm are discussed. They are too numerous to elaborate the details here, but we would like to point out about one addition (with respect to what was presented in [\onlinecite{cpc_219_407}]) which specifically was implemented in the course of our present study. Namely, the calculation of W for a specific momentum and Matsubara frequency was performed by a process group to distribute peak memory usage. This has been achieved by creating an interface with the ScaLAPACK library.\cite{scalapack} Our FLAPW+LO basis set consisted of approximately 1000 functions and included semi-core states (3s,3p of Co; 3s,3p,3d of As; 4s,4p,4d of Sb) as well as one additional high energy LO for s,p,and d orbitals of all atoms. Green's function and self energy (fermionic functions) in the LQSGW part of the calculations were expanded using all band states generated from the above FLAPW+LO basis set. In this way, the well known issue of slow convergence of GW-based methods with respect to high energy states was properly addressed. The size of the Product Basis (PB) for the bosonic functions (polarizability and screened interaction) slightly exceeded 4000, with approximately equal number of functions from the muffin-tin (MT) spheres and from the interstitial region. The radii of the MT spheres were 2.22 a.u. (Co and As), and 2.55 a.u (Sb). Inside the muffin-tin spheres the spherical harmonic expansion was terminated at $L_{max}=4$ for both the LAPW basis and the product basis. In all calculations we started with 80 iterations within GGA to seed subsequent LQSGW/hybrid iterations. After that, 20 iterations of LQSGW/hybrid functionals were sufficient to converge the bandgap better than 0.005eV.

Following the calculation of electronic structure, the temperature- and direction-dependent Seebeck coefficient was evaluated using Boltzman theory\cite{Boltzman}. The band energies obtained on the $8\times 8\times 8$ \textbf irreducible {k}-mesh were interpolated\cite{pr_178_1419,jcompp_67_253,prb_38_2721} onto finer $32\times 32\times 32$ \textbf{k}-mesh in the Brillouin zone. Comparison to control calculations with a $64\times 64\times 64$ \textbf{k}-mesh confirms that the results were sufficiently converged on the $32\times 32\times 32$ \textbf{k}-mesh. In the initial experimental study, Sb vacancies were observed but the doping level is unknown. The Seebeck coefficient was maximized (in absolute value) at 450 K $S_{max}$. Goldsmid and Sharp have shown that the temperature where the Seebeck coefficient maximizes ($T_{max}$) is a function of doping level, but their product, the Goldsmid-Sharp gap $E_{GS}=2*S_{max}T_{max}$ is invariant to the doping level. Therefore, all transport quantities were calculated at the doping level which produced $T_{max}=450 K$. The equations of the Boltzman theory for transport properties were also implemented as a part of the FlapwMBPT code. That was done (instead of using Boltztrap code) because our study deals essentially with contributions to transport properties beyond the leading bandgap-dependent terms including temperature-dependent shifts in the chemical potential. Also, taking into account possible future developments of the code, it was preferable to us to incorporate the corresponding models into FlapwMBPT.

We have used a model to address the uncertainty related to the non-stoichiometry of the compound, which we discuss here. Instead of changing the geometry, we assumed that impact of the non-stoichiometry on the transport properties can be modeled by considering the system as doped. Experimental information\cite{cm_30_4207} suggests that there should be an excess of electrons (n-type semiconductor). As it was discussed by Goldsmid and Sharp in [\onlinecite{jem_28_869}], the maximum in Seebeck coefficient as a function of temperature is formed where the excitation of extrinsic carriers is superceded by excitations of electron-holes pairs across the energy gap. Thus, in order to model the low temperature part (extrinsic carriers), some source of doping needs to be introduced. But for the modeling the high temperature regime, the host band structure is needed. Doping affects $T_{max}$, so for each method, the doping level was adjusted to achieve $T_{max}=450K$ which provides a consistent comparison between the electronic structure methods.

\section{Results and discussion}
\label{res}

\begin{figure}[t]
\centering
\includegraphics[width=8.0 cm]{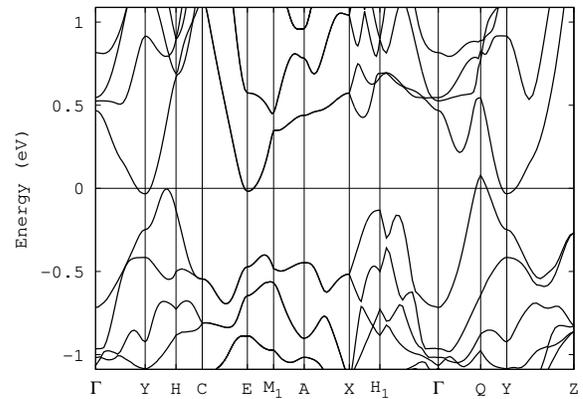}
\caption{Band structure of CoAsSb obtained in GGA approximation. The valence band has a maximum at point Q which is near the path between Y and H. The conduction band has minima at Y and E. Both the valence band and conduction band cross the Fermi level. } \label{b_o_gga}
\end{figure}

\begin{figure}[b]
\centering
\includegraphics[width=8.0 cm]{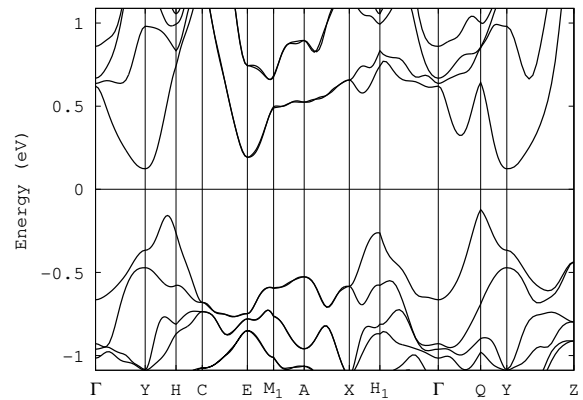}
\caption{Band structure of CoAsSb obtained in LQSGW$^{PBE}$ approximation. Fourier interpolation\cite{pr_178_1419,jcompp_67_253,prb_38_2721} has been used to obtain the band energies along the path in \textbf{k}-space. The valence band has a maximum at point Q which is near the path between Y and H. The conduction band has minima at Y and E.} \label{b_e_qp}
\end{figure}

\begin{table}[t]
\caption{Bandgap (eV, second column) and Goldsmid-Sharp gap, (eV, third column)\cite{jem_28_869}. For GGA, the band overlap is presented as a negative bandgap. Doping level at $T_{max} = 450 K$ (electrons per unit cell, fourth column). Maximum (most negative) Seebeck coefficient $S_{max}$ ($\mu V/K$, fifth column). Fractions of exact exchange for hybrid functionals are shown in brackets. Calculations using the scissor operator are marked with (s).} \label{e_dif}
\begin{center} 
\begin{tabular}{@{}c c c c c}
\hline \hline
Method & Bandgap & GS gap& Doping& $S_{max}$\\
\hline
GGA &-0.078 &&0.03&-74\\
Hybrid (0.1) &0.102 &0.155&0.009&-172\\
Hybrid (0.15) &0.230 &0.234&0.003&-260\\
Hybrid (0.25) &0.444 &0.396&0.0003&-440\\
LQSGW &0.279 &0.324&0.0032&-360\\
LQSGW$^{PBE}$ &0.246 &0.288&0.006&-320\\
GGA(s) &0.12 &0.188&0.015&-209\\
Hybrid (0.25)(s) &0.12 &0.155&0.007&-172\\
LQSGW (s) &0.12 &0.198&0.016&-220\\
LQSGW$^{PBE}$ (s) &0.12 &0.212&0.018&-235\\
Exp. &0.12-0.256 &0.121&&-134\\
\hline \hline
\end{tabular}
\end{center}
\end{table}

\begin{figure}[b]
\centering
\includegraphics[width=8.0 cm]{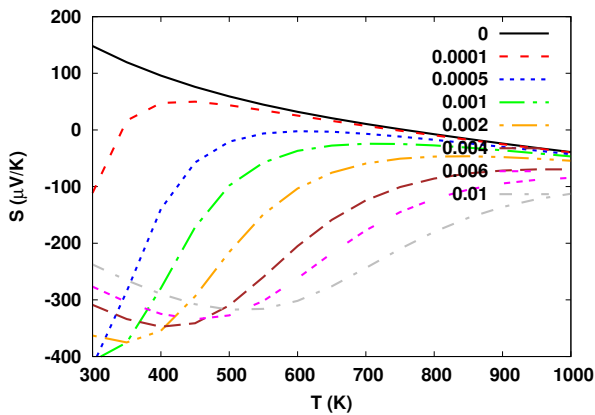}
\caption{Seebeck coefficient obtained with LQSGW$^{PBE}$ at different doping levels. 7 doping levels from 0.0001 to 0.01 electrons/unit cell along with zero doping are presented.} \label{seeb_74a}
\end{figure}

\begin{figure}[t]
\centering
\includegraphics[width=8.0 cm]{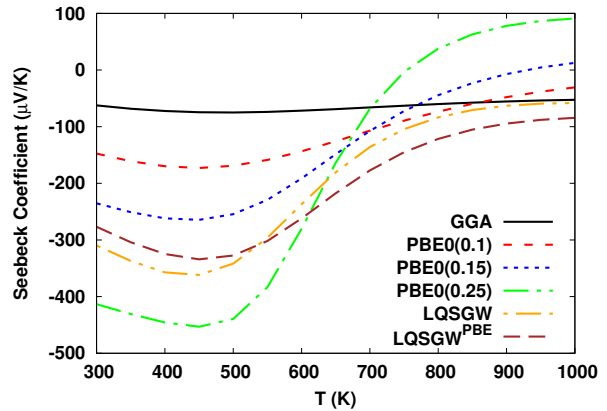}
\caption{Temperature-dependent Seebeck coefficient obtained using each method. Doping was applied to achieve a maximum Seebeck coefficient at $T_max = 450 K$.} \label{seeb_all}
\end{figure}

\begin{figure}[b]
\centering
\includegraphics[width=8.0 cm]{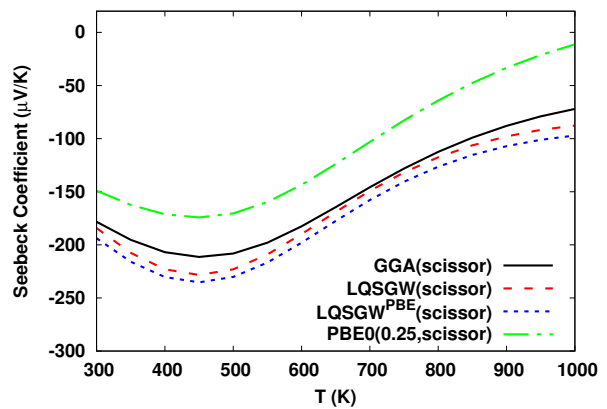}
\caption{Temperature-dependent Seebeck coefficient for each electronic structure method followed by scissor operator to achieve a bandgap of 0.12 eV. Doping was applied to achieve a maximum Seebeck coefficient at $T_max = 450 K$.} \label{seeb_scis}
\end{figure}

 Bandgaps obtained from each electronic structure method are shown in Table \ref{e_dif}. For the experimental geometry, GGA results in a metal (Figure \ref{b_o_gga}) in agreement with work [\onlinecite{cm_30_4207}]. LQSGW results in a band gap of 0.279 eV which is slightly above the experimental range 0.12-0.256 eV.\cite{cm_30_4207} The range of exact exchange in PBE0 (10\%-25\%) results in bandgaps which cover the experimental uncertainty. For the optimized geometry, we obtained a tiny gap in our DFT calculation (0.005 eV), which can be attributed to using GGA instead of LDA as in [\onlinecite{cm_30_4207}]. But with LQSGW, a noticeable overestimation of the band gap (0.441eV) is observed. We conclude that the DFT-optimized geometry of the stoichiometric compound gave a worse estimation of the band gap than the experimental geometry, and thus we will adhere to the experimental geometry henceforth.

To elicit the differences between the electronic structures computed by GGA and $LQSGW^{PBE}$, their band structures are compared. The band structure from GGA is shown in Figure \ref{b_o_gga} whilst Figure \ref{b_e_qp} shows the band structure from LQSGW$^{PBE}$. Notable differences are observed at some high symmetry points which reshape band edges or shift them more at some wave vectors than others. As it follows from the Table \ref{e_dif}, LQSGW$^{PBE}$ results in a better band gap as compared to LQSGW. Thus, we conclude that vertex corrections are essential and further improvement in the calculated band structure can be achieved with advanced diagrammatic approaches. 

We now turn to the consideration of thermoelectric properties. The Seebeck coefficient calculated with LQSGW$^{PBE}$ approach is presented in Figure \ref{seeb_74a} for various doping levels. Zero doping results in a positive Seebeck coefficient at low temperatures and a negative Seebeck coefficient at high temperatures in contrast to the consistent negative Seebeck coefficient from the experiment(Figure 7a in [\onlinecite{cm_30_4207}]). From a small increase in the doping, the LQSGW$^{PBE}$ curves suddenly become much more similar to the experiment and $T_{max}$ increases. The correct position of the maximum is achieved at the doping 0.006 electrons per unit cell which we compare to the experimental composition $CoAsSb_{0.883}$. The value of the Seebeck coefficient at the maximum ($S_{max}$) is -320 $\mu V/K$ which exceeds (in amplitude) the experimental value -134 $\mu V/K$\cite{cm_30_4207} reflecting the overestimation of the calculated band gap. For each method and the experiment, $S_{max}$ and the GS gap are collected in Table \ref{e_dif} where $T_{max} = 450 K$. Interestingly, the experimental GS gap deduced from the Seebeck coefficient (0.121 eV) coincides with bandgap deduced from conductivity from 325-475 K (0.12eV). The bandgaps and the GS gap are quite similar to one another, however the deviations among methods are intriguing. For LQSGW and $LQSGW^{PBE}$, the GS gap is greater than the bandgap. For the hybrid functionals, the GS gaps are greater than the lowest bandgap hybrid (bandgap: 0.102 eV, GS gap: 0.155 eV) but lower than the highest bandgap hybrid (bandgap: 0.444 eV, GS gap: 0.396 eV). The doping level necessary to obtain $T_{max} = 450 K$ mostly decreases as the bandgap increases. When the scissor operator was applied, the GS gap was always significantly above the bandgap. 

Figure \ref{seeb_all} presents Seebeck coefficient calculated as a function of temperature for each method used in this study and it is compared with the experimental function (Fig. 7a in Ref.[\onlinecite{cm_30_4207}]). Experimental values of S are: 300 K( -100 $\mu V/K$), 450 K( -134 $\mu V/K$) and 1000K (-75 $\mu V/K$). The overall ladle shape is reproduced by each method except GGA. Each method shows a smaller magnitude at high temperatures and each is less than 100 $\mu V/K$ in magnitude, however each method behaves somewhat differently at low temperature. For the hybrid functionals, the larger the gap, the larger the difference between low and high temperature coefficients and the 3 cross each other. Interestingly, although the hybrid functional  which corresponds to the best calculated band gap (0.1 of exact exchange) is somewhat more negative than the experiment at 300K and 450 K, at higher temperature it is more positive than the experiment. The more modest changes in temperature found for LQSGW and $LQSGW^{PBE}$ suggest that the dynamical correlation effects which are missing in the hybrid functionals bring important features in the spectra. The large deviation from experiment at low temperatures can be understood from the fact that extrinsic carriers are the dominant contributor here and correctly assigning extrinsic carriers has proven difficult. 

For GGA, PBE(0.25), LQSGW and $LQSGW^{PBE}$, we also studied the effect of a rigid shift of the conduction bands (scissor operator) which results in the experimental band gap - an approach which is quite popular when using DFT. Figure \ref{seeb_scis} presents the results. The curves are mostly parallel to one another and have similar temperature dependence. Notably, GGA gained the curved shape of other methods and more closely matches the experimental temperature dependence, but not the magnitude of the Seebeck coefficient. We find that this justifies the application of scissor operator to the DFT band structure. We note that the PBE(0.25) curve retained much of its large swing with temperature. We think, that this fact reflects the above mentioned differences in the shape of the bands, which is a testimony of the importance of dynamic correlations in studies of this material and thermoelectrics in general.

\section*{Conclusions}
\label{concl}

In conclusion, we have extended the theoretical study of CoAsSb performed in Ref. [\onlinecite{cm_30_4207}] by applying advanced (as compared to DFT) electronic structure methods LQSGW and  LQSGW$^{PBE}$ along with semiempirical hybrid functional (PBE0) with a few different percentages of the exact exchange. The bandgaps from LQSGW (0.279 eV) and $LQSGW^{PBE}$ (0.246) are a significant improvement from the gapless DFT calculations. If anything the bandgap is now too large, however this is complicated by apparent differences in the experimental gap at low temperature (0.12 eV) and high temperature (0.256 eV). Boltzman semiclassical transport theory was used to estimate the Seebeck coefficient from electronic structure. A full estimation of temperature-dependence was performed by using an extrinsic carrier density which reproduced the temperature of the Seebeck coefficient maximum. Additionally, the temperature-dependent Seebeck coefficient is significantly improved especially at high temperature. The effect of the scissor operator on the Seebeck coefficient was compared to advanced electronic structure methods by equating the bandgaps. This resulted in excellent agreement between GGA, LQSGW, and $LQSGW^{PBE}$, however all 3 resulted in a Goldsmid-Sharp gap much larger than the experiment. The remaining differences between the experimentally-measured and calculated Seebeck coefficient can primarily be ascribed to i) difficulty in assigning extrinsic carrier density given the non-stoichiometry found in the experiment and ii) absence of explicit diagrammatic vertex corrections in our calculations. Which one of the two source is more important is not clear at this point. Possible future extension of this work (by including diagrammatic vertex corrections) might allow to reduce the uncertainty in this respect.

\section*{Acknowledgments}
\label{ackn}

This work was   supported by the U.S. Department of Energy, Office of Science, Basic
Energy Sciences as a part of the Computational Materials Science Program.
A.R. thanks support from a NASA Space Technology Research Fellowship. We thank Chang-Jong Kang for sharing data on CoAsSb with us.


%

\end{document}